\documentclass[12pt]{iopart}

\usepackage{iopams}  
\usepackage{graphicx}
\usepackage{xcolor}

\begin{document}

\title[Whistler growth]{Weibel- and non-resonant Whistler wave growth in an expanding plasma in a 1D simulation geometry}

\author{M.~E.~Dieckmann}
\address{Dept. of Science and Technology (ITN), Link\"oping University, Campus Norrk\"oping, SE-60174 Norrk\"oping, Sweden}
\ead{mark.e.dieckmann@liu.se}

\author{L.~Palodhi}
\address{Department of Mathematics, Indian Institute of Technology Ropar, 140001, Punjab, India}

\author{C.~Fegan}
\address{Centre for Light-Matter Interactions, School of Mathematics and Physics, The Queen's University of Belfast, University Road, BT7 1NN, Belfast, United Kingdom}

\author{M.~Borghesi}
\address{Centre for Light-Matter Interactions, School of Mathematics and Physics, The Queen's University of Belfast, University Road, BT7 1NN, Belfast, United Kingdom}

\vspace{10pt}
\begin{indented}
\item[]August 2022
\end{indented}

\begin{abstract}
Ablating a target with an ultraintense laser pulse can create a cloud of collisionless plasma. A density ramp forms, in which the plasma density decreases and the ion's mean speed increases with distance from the plasma source. Its width increases with time. Electrons lose energy in the ion's expansion direction, which gives them a temperature anisotropy. We study with one-dimensional particle-in-cell simulations the expansion of a dense plasma into a dilute one, yielding a density ramp similar to that in laser-plasma experiments and a thermal-anisotropy-driven instability. Non-propagating Weibel-type wave modes grow in the simulation with no initial magnetic field. Their magnetic field diffuses across the shock and expands upstream. Circularly polarized propagating Whistler waves grow in a second simulation, in which a magnetic field is aligned with the ion expansion direction. Both wave modes are driven by non-resonant instabilities, they have similar exponential growth rates, and they can leave the density ramp and expand into the dilute plasma. Their large magnetic amplitude should make them detectable in experimental settings.  
\end{abstract}

%
\vspace{2pc}

\submitto{\PS}
%
\maketitle
%
%

\section{Introduction}

The absorption of an ultra-energetic laser pulse by a solid target heats its electrons to relativistic temperatures. Some electrons escape from the target surface and leave behind a positive net charge. An electric field develops between the sheath of escaped electrons and the positively charged surface, which accelerates its ions~\cite{Snavely2000,Wilks2001}. The ablated surface ions expand away from the target and the loss of surface ions gives rise to a rarefaction wave that propagates into the target. 

An ambipolar electric field grows in the density ramp between the dense and dilute plasma. It reaches a steady state once the net electron flow, which is caused by the thermal diffusion of electrons in the direction of the dilute plasma, is canceled by the electron flow due to the electric field. Ions are accelerated in the direction of the dilute plasma for as long as there is a plasma density change. In time, the ensemble of expanding ions changes into a beam with a mean speed that increases and a density that decreases with an increasing distance from the ablated target. We can understand the ion expansion as a hydrodynamic expansion of a blast shell into vacuum~\cite{Sack1987}, where the thermal pressure change is mediated by the ambipolar electric field rather than by binary collisions between gas particles. This electric field couples electron thermal energy to the expanding ions. If the plasma is collisionless, the electrons lose thermal energy in the direction of the electric field but not orthogonal to it since there is no force component along these directions; a temperature anisotropy develops. Electrons are cooler along the expansion direction than in the plane orthogonal to it. 

Weibel~\cite{Weibel1959,Romanov2004} showed that a magneto-wave grows in a spatially uniform plasma with such a thermal anisotropy. Its wavevector is aligned with the cool direction and its magnetic field is oriented in the plane orthogonal to it. This instability works as follows. Counterstreaming electrons repel each other magnetically while those moving in the same direction attract each other. This attraction leads to the formation of current channels filled with co-moving electrons that are separated magnetically from the current channel that contains electrons moving the other way. If the electron temperature is low in one direction, then the random thermal motion of electrons cannot counteract the formation of current channels in the plane orthogonal to this direction. 

Particle-in-cell (PIC) simulations have demonstrated that the Weibel instability grows in circular rarefaction waves~\cite{Dieckmann2012, Nechaev2020} and in planar rarefaction waves that expand into vacuum~\cite{Thaury2010} or into an ambient plasma~\cite{Fox2018}. The Weibel instability in rarefaction waves has been found in laser-plasma experiments, where a wire was ablated by an energetic laser pulse and where the blast shell temperature was high enough to let the plasma be collisionless~\cite{Quinn2012}. If the plasma is initially unmagnetized, the Weibel instability competes with the thermomagnetic instability~\cite{Tidman1974}, which grows wherever the temperature gradient is not aligned with the density gradient. 

A magnetic field with a sizeable amplitude changes the wave spectrum in the plasma, which enables the growth of waves other than the Weibel mode. We address the blast shell expansion's initial phase and focus on a collisionless instability with a large exponential growth rate; the Whistler wave instability. Whistler waves (See~\cite{Zhao2017} for a recent review) are widely observed in the near-Earth plasma and laboratory~\cite{Stenzel1999} for example in the Van Allen belt. Electrons in this belt bounce back and forth between the Earth's magnetic poles and the conservation of magnetic momentum in the converging magnetic field lines near the poles reflects most of them before they reach the upper atmosphere. However, only electrons with a large pitch angle relative to the magnetic field are reflected. A loss-cone distribution forms with a larger thermal energy perpendicular to the magnetic field than parallel to it, which leads to the growth of Whistler waves by their resonance with the gyrating electrons~\cite{Dungey1963,Devine1995,Millan2007}. Another wave growth mechanism is an electric field, which points along a background magnetic field and pushes them into a cyclotron resonance with the Whistler waves~\cite{Das1982}. 

In this paper, we want to determine if, where, and how Whistler waves can grow in the plasma of a blast shell, which expands into an ambient plasma. Like in Ref.~\cite{Dieckmann2010}, we place a dense plasma into the center of a one-dimensional simulation box and surround it with a low-density plasma. We compare the results of a simulation without a background magnetic field with those of a simulation where we align a magnetic field with the simulation direction. In both simulations, rarefaction waves propagate into the dense plasma starting at both boundaries. Ions are accelerated in the opposite direction. Shocks form between the moving ions and the ambient ions at rest. Ions are accelerated at the expense of the electron's thermal energy in the simulation direction and the electron distribution becomes thermally anisotropic in the density ramps between the dense plasma and the shocks. The density ramps widen in time and eventually, the electron distribution becomes unstable. A Weibel-type instability leads to the growth of non-propagating magneto-waves in the simulation with no background magnetic field. Their magnetic field diffuses across the shock into its foreshock region. In the simulation with the magnetized plasma, the magneto-waves propagate out of the density ramp. We identify them as Whistler waves based on their approximately circular polarization and their large propagation speed. Given that these magneto-waves grow and saturate on a time scale that is comparable to an inverse electron gyrofrequency and that their exponential growth rate is close to that of the Weibel-type modes, we propose that their growth mechanism is the non-resonant Whistler wave instability~\cite{Lazar2009,Palodhi2010} where waves are driven by a thermal anisotropy of the electrons as a whole like it is the case for the Weibel instability~\cite{Weibel1959,Stockem2009,Treumann2018}. 

Our paper is structured as follows. Section~2 discusses relevant aspects of Whistler waves, the PIC code, and the initial conditions we used for our simulations. Section~3 presents our results, which we summarize in Section~4.

\section{Whistler waves, the PIC code, and the initial conditions}\label{sec:introwaves}

Consider a spatially uniform plasma, which consists of cold electrons and is permeated by a magnetic field $\mathbf{B}_0$ with $B_0=|\mathbf{B}_0|$. Ions form a positively charged immobile background. The electron number density $n_0$, mass $m_e$, and charge $-e$ gives the plasma frequency $\omega_p={(n_0e^2/m_e\epsilon_0)}^{1/2}$ ($e,\epsilon_0$: elementary charge and vacuum permittivity). The electron gyro-frequency is $\omega_{ce}=eB_0/m_e$. Whistler modes in cold electron plasma are electromagnetic waves with a right-handed polarization. If their wave vector $\mathbf{k}$ points along $\mathbf{B}_0$, they are circularly polarized, they propagate along $\mathbf{B}_0$, and follow the dispersion relation 
\begin{equation}
\omega (k) = \frac{\omega_{ce}}{1+(\omega_{p}/kc)^2},
\label{eq1}
\end{equation}
where $\omega, k$ are the wave's frequency and wave number along $\mathbf{B}_0$.

Whistler waves can resonate with electrons, which gyrate in the same direction. Consider an electron, which moves at the speed $v_0$ along $\mathbf{B}_0$, and a monochromatic wave that propagates with the phase speed $v_{ph}=\omega / k$ in the opposite direction. In the reference frame moving with the electron speed $v_0$, the Doppler-shifted wave frequency is $\tilde{\omega} = (1 + \frac{v_0}{v_{ph}}) \omega = \omega + v_0k$ and the condition for resonance with the electron is $\tilde{\omega}=\omega_{ce}$. We obtain their resonance for $\omega - \omega_{ce} = -v_0k$. 
Hot electrons in thermal equilibrium have a Maxwellian velocity distribution along $\mathbf{B}_0$ with the thermal speed $v_{th,e}={(k_BT_e/m_e)}^{1/2}$ ($k_B, T_e$: Boltzmann constant and electron temperature). Most electrons move with $-v_{th,e}\lesssim v_0 \lesssim v_{th,e}$ along $\mathbf{B}_0$ and each can resonate with the Whistler wave. For $k\ge 0$, we expect strong wave-particle interactions for
\begin{equation}
|\omega - \omega_{ce}|/ v_{th,e} \lesssim k.
\label{eq2}
\end{equation}
 
According to Eqn.~\ref{eq1}, $\omega$ goes to zero in the limit $k\rightarrow 0$ and such Whistler waves cannot interact resonantly with electrons. With increasing $k$, the frequency $\omega$ approaches $\omega_{ce}$ while the range of resonant frequencies set by Eqn.~\ref{eq2} broadens. Whistler waves with a short wavelength can resonate with electrons. Resonant interactions are efficient if the electron remains in phase with the wave for longer than $\omega_{ce}^{-1}$. Resonant interactions between electrons and Whistler waves damp the modes if the electrons are in thermal equilibrium and waves can grow otherwise. 

Particle-in-cell (PIC) codes can resolve Whistler waves and their interactions with electrons. We use the EPOCH code, which solves Maxwell's equations on a grid. It evolves the electric field $\mathbf{E}(x,t)$ and magnetic field $\mathbf{B}(x,t)$ in time using Amp\`ere's law and Faraday's law 
\begin{equation}
\nabla \times \mathbf{B} = \mu_0 \mathbf{J} + \mu_0 \epsilon_0 \frac{\partial \mathbf{E}}{\partial t}, \hskip .5cm
\nabla \times \mathbf{E} = -\frac{\partial \mathbf{B}}{\partial t}.
\label{eq3}
\end{equation}
Esirkepov's scheme~\cite{Esirkepov2001} fulfills Gauss' law $\nabla \cdot \mathbf{E} = \rho / \epsilon_0$ ($\rho$: charge density) and $\nabla \cdot \mathbf{B}=0$ as constraints. The plasma is represented by an ensemble of computational particles (CPs), which have the same charge-to-mass ratio as the species they represent. The current contributions of all CPs are interpolated from their position to the numerical grid and summed up to give the macroscopic current density $\mathbf{J}(x,t)$, which is used by Amp\`ere's law to update the electromagnetic fields in time. The updated electromagnetic fields are interpolated back to the position of each CP $i$ and its momentum is updated with the numerical approximation of the Lorentz force equation $\dot{\mathbf{p}}_i =q_i (\mathbf{E}(\mathbf{x}_i)+\mathbf{v}_i \times \mathbf{B}(\mathbf{x}_i))$ ($q_i,\mathbf{x}_i,\mathbf{v}_i,\mathbf{p}_i$: charge, position, velocity and momentum of $i^{th}$ CP). See~\cite{Arber2015} for an in-depth description of the numerical scheme. All simulations discussed here resolve one spatial $x$ direction, all three particle velocity components, all three components of $\mathbf{E}$ and $\mathbf{B}$, and use periodic boundary conditions. 

In the next section, we perform two simulations, which track the expansion of a dense plasma into an ambient plasma. The first simulation considers an initially unmagnetized plasma. In the second simulation, we align a background magnetic field with the simulation direction $x$ and set $\omega_{ce}=0.084 \omega_p$. Our ambient plasma consists of electrons and fully ionized nitrogen ions with the mass $m_i = 2.57\times 10^4m_e$. Nitrogen is widely used as residual gas in laser-plasma experiments and is turned into ambient plasma by ionizing secondary X-ray radiation from the ablated target~\cite{Ahmed2013}. The mean velocity of both particle species is zero at the simulation's start and the plasma is thus charge- and current-neutral at the time $t=0$. Initially, the velocity distributions of the ambient plasma's electrons and ions are isotropic in velocity space and their temperatures are $T_e=10^3$ eV and $T_i = T_e/5$, respectively. Before we discuss the results of both simulations, we explore some properties of the magnetized ambient plasma with a separate PIC simulation that neither contains the dense plasma nor mobile ions.

In a PIC simulation, each CP contributes a micro-current that is proportional to its speed and charge. It induces an electromagnetic field pulse with a phase speed that matches the speed of the CP along the simulation direction. Typically, this perturbation is not an undamped eigenmode of the plasma, and the wave energy gets reabsorbed by other CPs. In time an equilibrium is established between wave emission and absorption. Fourier transforming these fields over space and time reveals the range of $\omega,k$ where waves can interact with electrons. Statistical noise in PIC simulations is closely related to thermal noise in plasma as has been demonstrated for unmagnetized plasma~\cite{Dieckmann2004}. We will use noise to gain insight into the wave properties of the ambient plasma. 

\begin{figure}
\includegraphics[width=\columnwidth]{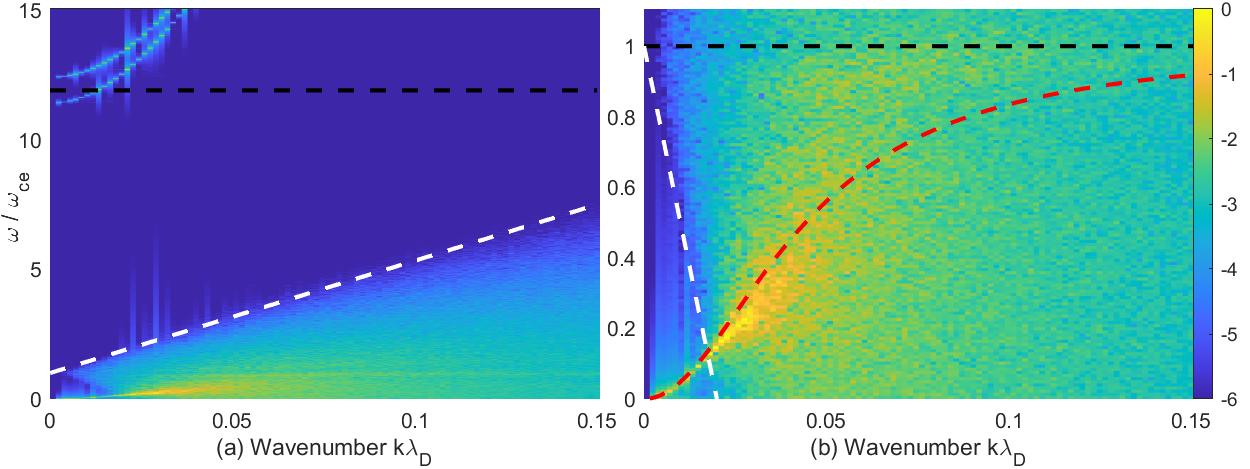}
\caption{The power spectrum $P_B(k,\omega) = |B_y^2(k,\omega)|+|B_z^2(k,\omega)|$ normalized to its peak value and displayed on a 10-logarithmic color scale. Panel~(a) shows the power spectrum up to frequencies beyond the electron plasma frequency $\omega_p/\omega_{ce}=11.9$ (dashed black line). The dashed white line is $\omega_{r+}=1+50k$. Panel~(b) zooms in on the low-frequency range. The dashed black line and the dashed red curve are $\omega_{ce}$ and the dispersion relation of Whistler waves. The dashed white line is $\omega_{r-}(k)=1-50k$.}
\label{figure1}
\end{figure}

We normalize time to $\omega_{ce}^{-1}$ and space to $\lambda_D$, where $\lambda_D = v_{th,e}/\omega_p$ is the Debye length of electrons with the density $n_0$ and temperature $T_e$. We take a simulation box with the length $L_x = 4000$ along $x$, which we resolve by 20000 grid cells. Electrons are represented by 500 CPs per cell. The simulation resolves the time interval $0 \le t \le 1420$ by $8\times 10^5$ time steps. Current density fluctuations orthogonal to $x$ give rise to fluctuations in the magnetic field. We write out the magnetic $B_y(x,t)$ and $B_z(x,t)$ components and split each of them into two segments. One covers $0 < t\le 710$ and the second $710 < t \le 1420$. Each segment is Fourier-transformed over space and time and multiplied with its complex conjugate. These four power spectra are summed up to $P_B(k,\omega)=|B_y^2(k,\omega)|+|B_z^2(k,\omega)|$, where the frequency $\omega$ and wave number $k$ are normalized to $\omega_{ce}$ and $\lambda_D^{-1}$, and shown in Figure~\ref{figure1}.

Figure~\ref{figure1}(a) covers frequencies beyond $\omega_p / \omega_{ce} = 0.084^{-1}$. The two high-frequency branches near $\omega_p$ are the left-handed and right-handed high-frequency modes. In the limit $B_0 \rightarrow 0$, both converge to the ordinary mode. Figure~\ref{figure1}(b) shows low frequencies. The red dashed curve marks the solution of the dispersion relation Eqn.~\ref{eq1}. A broad noise peak is enclosed by the fitted dashed white lines $\omega_{r+}(k) = 1 + 4.2 (\omega_p/\omega_{ce})k = 1+50k$ and $\omega_{r-}(k) = 1-50k$. In physical units, both lines enclose
\begin{equation}
|\omega-\omega_{ce}|/4.2v_{th,e} \le k.
\label{eq4}
\end{equation}
The factor of 4.2, which is not present in Eqn.~\ref{eq2}, shows that there are enough CPs with speeds $-4.2 v_{th,e} \le v_x \le 4.2 v_{th,e}$ along $x$ that can resonate with electromagnetic waves. According to Figure~\ref{figure1}(b), the noise power is enhanced near the solution of the Whistler mode's dispersion relation. Up to the wave number $k\approx 0.02$, the low-frequency noise power peaks close to the dispersion relation of the Whistler mode. Electrons interact via cyclotron resonance with the Whistler modes for $k>0.02$. The noise distribution broadens in the $\omega$-direction and eventually vanishes for $k > 0.05$. Whistler waves with $k<0.02$ do not undergo cyclotron resonance with the electrons for the temperature $T_e$ and for the number of CPs we use. This wavenumber corresponds to a wavelength $\sim 300$. The waves interact with energetic electrons for $0.02 \le k \le 0.05$ with the latter $k$ corresponding to a wavelength $\sim 120$.

\section{Simulations of expanding plasma}

In what follows, we normalize space to $\lambda_D$ and time to $\omega_p^{-1}$ of the ambient plasma. Initially, the dense plasma is located in the interval $-107 \le x \le 107$. Its density is 60 times that of the ambient plasma and its initial electron temperature is $1.5T_e$. The higher electron temperature accounts for the positive electric potential the dense plasma has relative to the ambient one. The electric field is expressed in units of $m_e c \omega_{p}/e$ and unless stated otherwise the magnetic field in units of $m_e \omega_{p}/e$. In this normalization, $B_0 = 0.084$. Ion densities $n_i(x,t)$ are normalized to the initial ion density of the ambient plasma $n_0/7$. Ion velocities are normalized to the ion acoustic speed $c_s=(k_B(\gamma_e Z T_e + \gamma_i T_i)/m_i)^{1/2}$, where $\gamma_e$ and $\gamma_i$ are the adiabatic constants of the electrons and ions. On ion time scales, electrons have enough time to diffuse in phase space due to scattering by electromagnetic field fluctuations. Therefore they have three degrees of freedom in unmagnetized plasma giving $\gamma_e = 5/3$. Ions are accelerated by the unidirectional electric field of the ion acoustic wave and have one degree of freedom giving $\gamma_i = 3$. In fully ionized nitrogen with $Z=7$, we obtain the value $c_s\approx 2.8 \times 10^5$ m/s or $v_{th,e}/c_s\approx 47$.

\subsection{Unmagnetized plasma}

\begin{figure}[ht]
\includegraphics[width=\columnwidth]{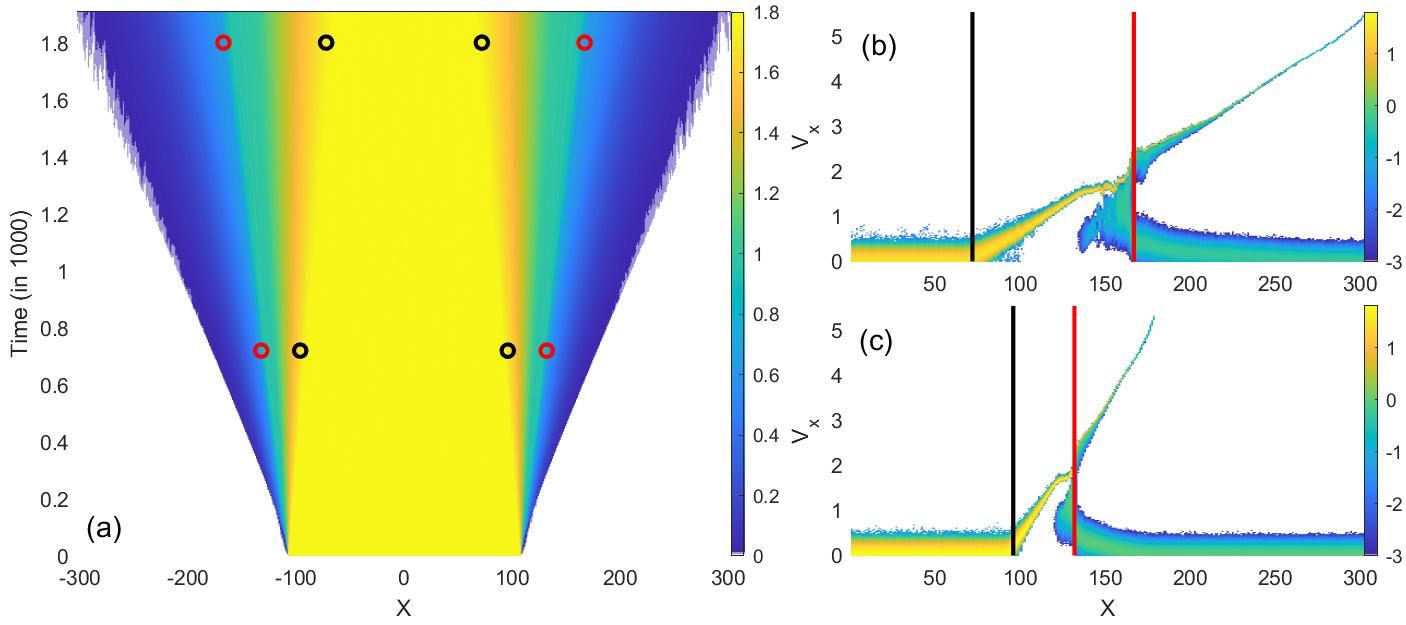}
\caption{The ion expansion: Panel~(a) shows the 10-logarithm of $n_i(x,t)$. The red circles show the points $(x,t)=(\pm 132, 720)$ and $(\pm 167,1800)$. The black circles show the points $(\pm 96, 720)$ and $(\pm 72, 1800)$. Panel~(b) displays the 10-logarithmic ion phase space density distribution for $x>0$ and $t=1800$ and panel~(c) that at $t=720$. The phase space density is normalized to its peak value. The black and red lines correspond to the circles at the same positions and times in~(a).}
\label{figure02}
\end{figure}

Figure~\ref{figure02} shows the ion expansion during the full simulation time.
Initially, the dense plasma is located in the interval $|x| \le 107$. Two rarefaction wavefronts propagate into the dense plasma with the speed $c_s$ and accelerate the ions outwards. Both are indicated by the black circles at $t=720$ and $t=1800$ in Fig.~\ref{figure02}(a). Figures~\ref{figure02}(b,~c) reveal that the rarefaction wavefronts separate ions of the dense plasma at rest from accelerating ones. The accelerating ions push those of the ambient plasma and a shock develops between the moving ions and the ions of the ambient plasma at rest. Figure~\ref{figure02}(a) demonstrates that two shocks, the positions of which are indicated by red circles at $t=720$ and $t=1800$, move away from the center of the dense plasma and into the ambient plasma. Based on the positions and times of the red circles, we obtain the shock speed $1.5c_s$. 

In what follows, we refer with thermal dense plasma to the interval to the left of the black lines in Figs.~\ref{figure02}(b,~c), where ions are at rest. The density ramp is located to the right of the black line where a single ion population is located. The shock's downstream region is located to the left of the red line where we find a hot population of accelerated ambient ions and a cool beam of fast ions from the dense plasma. In Fig.~\ref{figure02}(b), the boundary between the density ramp and the downstream is located at $x\approx 135$. To the right of the red line, the foreshock is the interval where the ion distribution is not thermal. Thermal ambient plasma is located in the interval $x>180$ in Fig.~\ref{figure02}(c). 

\begin{figure}[ht]
\includegraphics[width=\columnwidth]{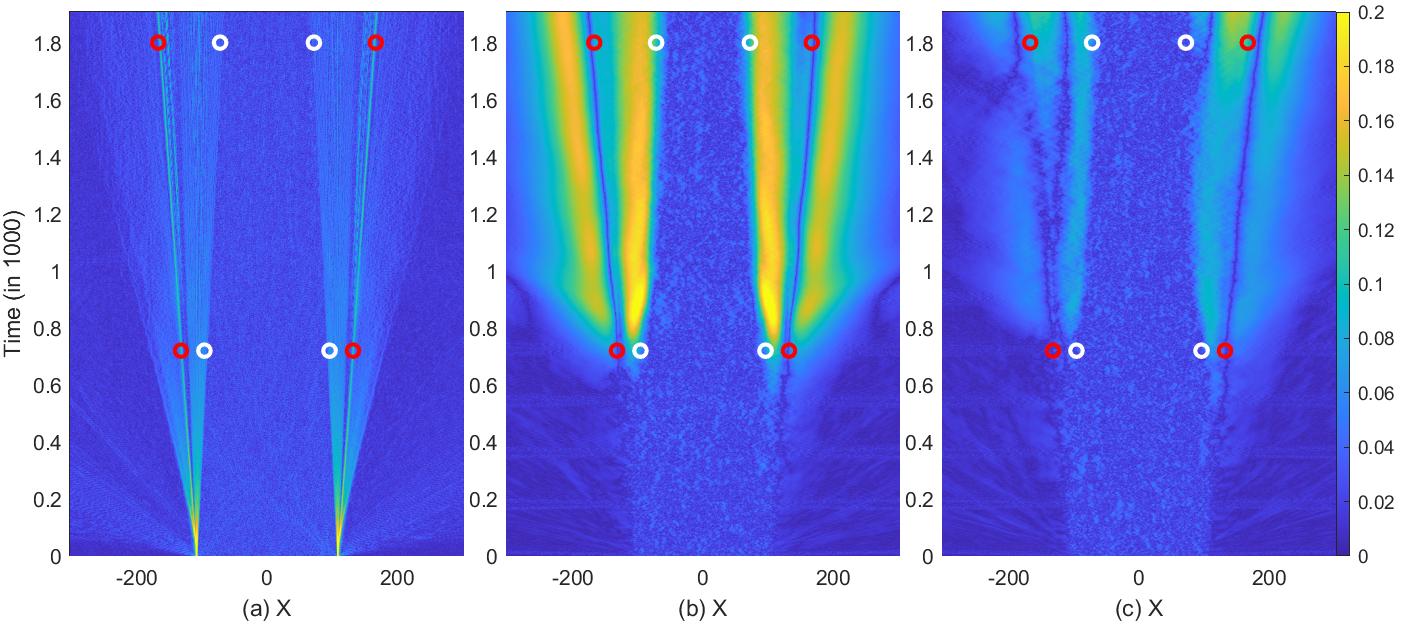}
\caption{Time evolution of the electromagnetic fields: Panel~(a-c) show $|E_x(x,t)|^{1/2}$, $|B_y(x,t)|^{1/2}$, and $|B_z(x,t)|^{1/2}$. Red circles show the positions of the shocks and the white ones are those of the fronts of the rarefaction waves at $t=720$ and $1800$. All panels use identical color scales.}
\label{figure03}
\end{figure}

Figure~\ref{figure03} shows the distributions of $|E_x(x,t)|^{1/2}$, $|B_y(x,t)|^{1/2}$, and $|B_z(x,t)|^{1/2}$.
The strongest electric field peaks at the shock locations are trailed by weaker ones, which are caused by ion-acoustic waves in the downstream regions. These electric field peaks are surrounded by diffuse electric field patches, which are tied to slow density changes and constitute the aforementioned ambipolar electric field. On average, the amplitude of the narrow electric field peaks and of the surrounding electric field patches is positive for $x >0$ and negative for $x<0$; they point in the direction of decreasing density. After $t\approx 600$ magnetic fields grow in the density ramp and in the foreshock. They are linearly polarized and almost aligned with $y$ until $t=1600$ when $B_z$ starts to grow. Their amplitudes in Figs.~\ref{figure03}(b,~c) go through minima at the shock locations. 

If the magnetic field growth is caused by the Weibel instability, it must be connected to a directional anisotropy of the electron's thermal energy. We examine the thermal energies of electrons along $x$, $y$, and $z$. We project the electron phase space density distributions onto the planes $(x,v_i)$ with $i$ = $x$, $y$, and $z$. First, we compute the mean velocities $\langle v_{i}(x) \rangle =n_e(x)^{-1}\int v_i f_e(x,v_i) dv_i$ with the electron's number density $n_e(x)$ and phase space density distribution $f_e(x,v_i)$. The thermal energy per electron is $T_i = n_e(x)^{-1}\int (v_i-\langle v_i(x) \rangle)^2 f_e(x,v_i)dv_i$. The electron velocity distribution is not always a Maxwellian and $T_i$ is therefore not the temperature in the strict sense. We quantify the thermal anisotropy with $A_y(x)=T_y(x)/T_x(x)$ and $A_z(x)=T_z(x)/T_x(x)$.

\begin{figure}[ht]
\includegraphics[width=\columnwidth]{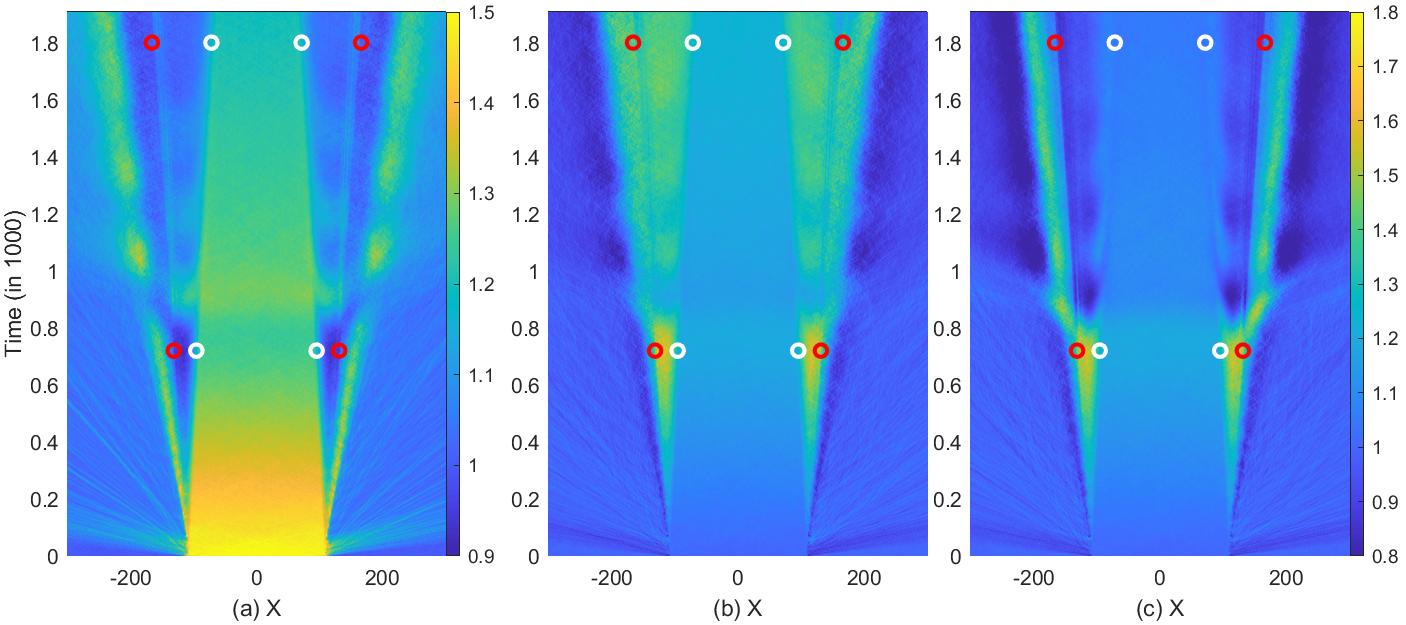}
\caption{The mean electron thermal energy $T_x(x)$ along $x$ is shown in panel~(a). It is normalized to that of electrons with the temperature $T_e$. Panels~(b,~c) show $A_y(x) = T_y(x)/T_x(x)$ and $A_z(x)=T_z(x)/T_x(x)$, respectively. Panels (b,~c) use the same color scale. Red circles show the positions of the shocks and the white ones are those of the fronts of the rarefaction waves at $t=720$ and $1800$.}
\label{figure04}
\end{figure}

Figure~\ref{figure04}(a) shows how $T_x(x)$ evolves from the initial distribution, when electrons in the dense and ambient plasma had the respective temperatures $1.5T_e$ and $T_e$, until the simulation's end $t=1900$. At the start of the simulation, hot electrons escape from the dense plasma while electrons from the ambient plasma are dragged into the dense plasma by the jump in the electric potential between both plasmas. Once the electron distribution in and near the dense plasma has reached an equilibrium, the electron temperature in the thermal dense plasma decreases and starts to oscillate at around $t=900$ when $B_y(x,t)$ saturates in Fig.~\ref{figure03}(b). The rate of change and the amplitude of the oscillations decrease over time. The oscillation period is around 250 time units. The maximum value $|B_y|^{1/2}=0.2$ in Fig.~\ref{figure03}(b) gives the electron gyrofrequency $\approx$ 0.04 or one gyroperiod $2\pi/0.04=160$, which is close to the observed one.

A thermal anisotropy develops in Figs.~\ref{figure04}(b,~c) in the density ramp and it peaks at $t\approx 720$. Its value decreases shortly after that and it starts to oscillate. This oscillation is tied to that of $T_x(x)$ and has the same frequency. Its decrease in the density ramp is minor in Fig.~\ref{figure04}(b) while the anisotropy practically disappears in the same region in Fig.~\ref{figure04}(c). A comparison with Fig.~\ref{figure03}(b,~c) shows that the anisotropy value decreases most in the direction perpendicular to that of the magnetic field. Electrons gyrate in the magnetic field, which depletes the anisotropy in the plane orthogonal to it. 

In time, the anisotropy value also increases in the thermal dense plasma. It keeps increasing in Fig.~\ref{figure04}(b) until the simulation's end while it decreases in Fig.~\ref{figure04}(c) after the Weibel-type modes reach their peak amplitude. Electrons that move from the thermal dense plasma into the density ramp are slowed down by the ambipolar electric field and some of them get reflected. This reflection is elastic because the dense plasma is expanding. The electrons are also rotated by the magnetic field, which yields the observed reduction of $A_z(x)$ after the magnetic field grows to a large amplitude. 

No Weibel-type modes grow in the thermal dense plasma. We can estimate the growth rate of the Weibel instability in this spatially uniform plasma based on Eqn.~5 in Ref.~\cite{Treumann2018} where we used our definition of the anisotropy value: $A_y(x)=1$ and $A_z(x)=1$ imply a stable electron distribution. The growth rate is
\begin{equation}
\sigma_W / \omega_p \approx {\left (\frac{8(A-1)}{27\pi} \right )}^{1/2}\left ( \frac{v_{th,e}}{c}\right ) A. 
\label{Weibel}
\end{equation}
The value of $\sigma_W$ in the thermal dense plasma with the local plasma frequency $\sqrt{60}\omega_p$ and the maximum value $A=1.15$ is $\sigma_W / \omega_p \approx 0.013$. This growth rate is probably too low to trigger a Weibel instability.

Before $t\approx 800$, the foreshock is populated by the hot electrons that escape from the dense plasma at $t\approx 100$. After $t\approx 800$, the value of $T(x)$ decreases and $A_y(x)$ and $A_z(x)$ increase in the foreshock. Once the anisotropy is established, a magnetic field grows (See Fig.~\ref{figure03}(b,~c)). It does not deplete the thermal anisotropy in the foreshock in Figs.~\ref{figure04}(b,~c), which suggests that the magnetic field is not growing due to a Weibel-type instability. The ambipolar electric field in the foreshock leads to a strong deviation of the electron velocity distribution along $x$ from a Maxwellian; a cool beam of ambient electrons is pulled into the downstream region of the shock while hot electrons flow upstream of the shock. Such a distribution may not be susceptible to the Weibel instability. We find an equally strong ambipolar electric field in the density ramp but the much larger plasma density implies that the effects on the phase space density distribution are smaller.

We may expect that in a system, which is symmetric around $x$, the growing magneto-waves are circularly polarized. We observe instead that the waves are linearly polarized and that the magnetic field is almost aligned with $y$.
The polarization of the growing waves is determined by noise. Random magnetic field fluctuations can, however, not explain why the magneto-waves have the same polarization on both sides of the dense plasma in Fig.~\ref{figure03}. We rearrange the magnetic field data into the complex-valued $B^*_\perp (x,t)=B_z(x,t)+iB_y(x,t)$ and compute its modulus $B_\perp(x,t)=|B^*_\perp (x,t)|$ and phase angle $\alpha (x,t)$. Selecting an imaginary $B_y(x,t)$ keeps the phase angles of the magneto-waves in this simulation in the interval $-\pi \le \alpha(x,t)\le \pi$. 
\begin{figure}[ht]
\includegraphics[width=\columnwidth]{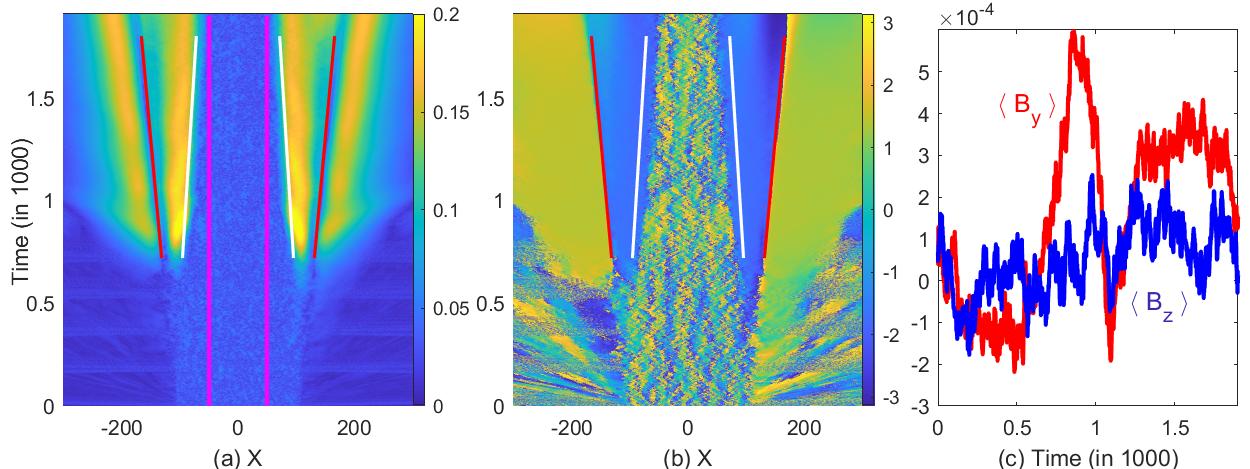}
\caption{The modulus/phase angle representation of the magnetic field. Panel~(a) shows $B_\perp^{1/2}(x,t)$. The vertical lines colored in magenta mark $|x|=50$. Panel~(b) shows the phase angle $\alpha(x,t)$ in radians. The lines connect the circles with the same color we previously plotted. Panel~(c) plots the magnetic components $\langle B_y \rangle$ and $\langle B_z \rangle$ that have been averaged over the interval enclosed by the lines $|x|=50$ in (a).}
\label{figure05}
\end{figure} 

The amplitude in Fig.~\ref{figure05}(a) goes through a minimum at the shock and remains low downstream of the shock. It increases with the plasma density in the ramp and decreases again as the thermal dense plasma is reached where it is shielded on a spatial scale $\sim 10$, which is a few times the electron skin depth in the dense plasma. By fitting $B_\perp(x=102,t)$ with an exponential function, we get the growth rate $\sigma_u \approx 0.14\omega_p$ of the instability noting that the local plasma frequency exceeds by far $\omega_p$ of the thermal ambient plasma. This growth rate is twice as high as the one we would get in a spatially uniform plasma from Eqn.~\ref{Weibel} with $A=1.8$ and a local plasma frequency $\sqrt{45}\omega_p$. 

Figure~\ref{figure05}(b) shows the phase angle between the magnetic field vector and the z-axis. The angle in the density ramp is about -1 while it is close to 2 in the foreshock on both sides of the dense plasma giving a rotation angle $\approx \pi$. The polarization of the magneto-waves on both sides of the thermal dense plasma is the same after $t=600$ but not at earlier times. In the density ramp near $x\approx -110$ and $t\approx 500$, the phase angle is about 0 while it is $-\pi$ at lower values of $x$. At early times, the magneto-waves were linearly polarized in both density ramps but their phase angles were not correlated as expected for waves that grew out of random noise.
 
Figure~\ref{figure05}(a) shows that the wave at $x\approx100$ started to grow before that at $x\approx -100$. It reaches $B_\perp(x,t)^{1/2}=0.08$ about 100 time units earlier. A signal propagating with the electron's thermal speed, which is higher in the thermal dense plasma than in the ambient one, could reach the other ramp during that time interval. Figure~\ref{figure05}(c) computes the mean value of the magnetic field components $\langle B_y \rangle$ and $\langle B_z \rangle$, which we averaged over the interval $|x|\le 50$ shown in Fig.~\ref{figure05}(a). During $500 \le t \le 600$, $\langle B_y \rangle$ grows and reaches a value that is well above noise levels and is also larger than $\langle B_z \rangle$. The magnetic field is shielded in the thermal dense plasma by an electric current that is orthogonal to $x$. The thermal motion of electrons along $x$ lets this current diffuse along $x$. It reaches the unstable region near $x=-100$ where it changes the orientation of the growing magnetic field. The electric current near $x\approx 100$, which confines the magnetic field in the direction of increasing $x$, spreads also to increasing $x$. Since $v_{th,e}/c_s = 47$ and because the shock's electric field along $x$ does not affect the electron speed along $z$, the electric current diffuses much faster along $x$ than the shock propagates. The electric current launched at the front of the magneto-wave at large $x$ decreases the magnetic amplitude with increasing $x$. If this electric current diffuses across the shock, it will induce a magnetic field in the foreshock that is rotated by $\pi$ relative to that in the density ramp. This rotation angle is observed at both shocks in Fig.~\ref{figure05}(b).

\subsection{Magnetized plasma}

\begin{figure}[ht]
\includegraphics[width=\columnwidth]{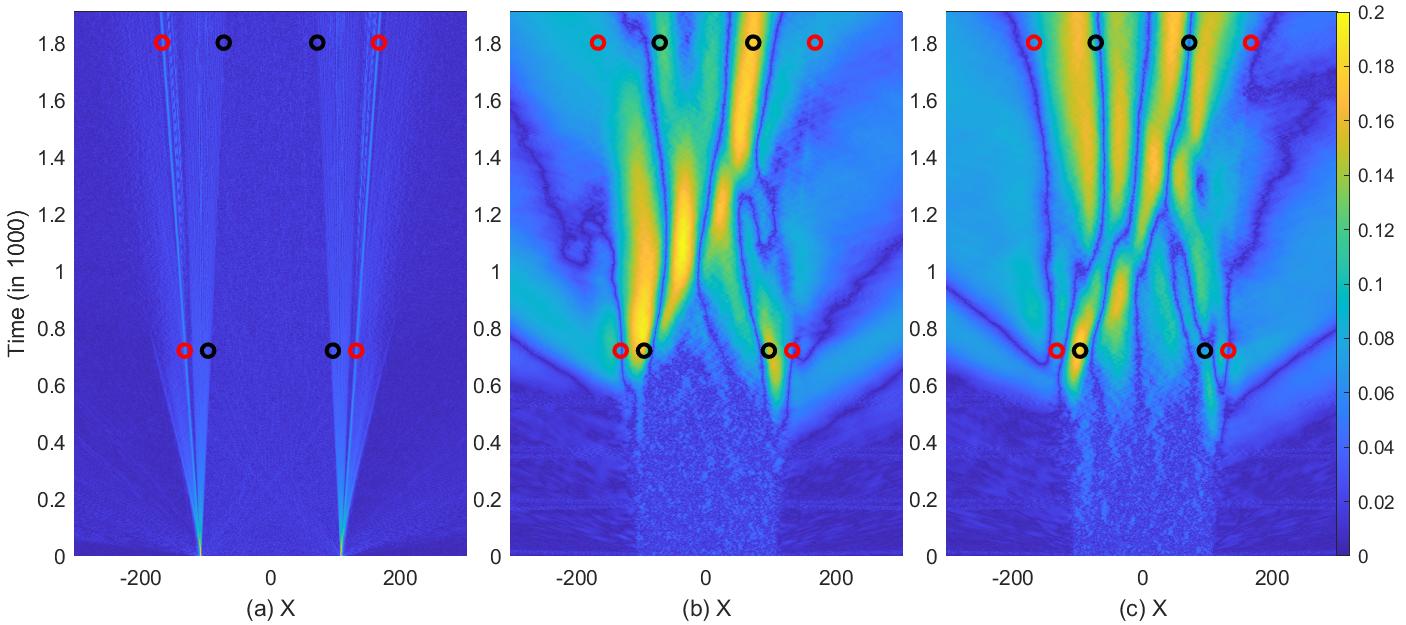}
\caption{Time evolution of the electromagnetic fields: Panels~(a-c) show $|E_x(x,t)|^{1/2}$, $|B_y(x,t)|^{1/2}$, and $|B_z(x,t)|^{1/2}$, respectively. All panels use the same color scale. The red circles show the locations of the shocks at $t=720$ and $t=1800$ while the black circles show those of the rarefaction wavefront. All panels use identical color scales.}
\label{figure06}
\end{figure}

Figure~\ref{figure06} shows $|E_x(x,t)|^{1/2}$, $|B_y(x,t)|^{1/2}$, and $|B_z(x,t)|^{1/2}$. That of $|E_x(x,t)|^{1/2}$ is practically identical to that in Fig.~\ref{figure03}(a). Electrons gyrate once around the background magnetic field $B_0=0.084$ during $\Delta_t = 2\pi / \omega_{ce}\approx 75$. The much heavier ions do not react to $B_0$ during the short time that is resolved by our simulation and electrons can stream freely along the magnetic field. The distributions of the ions and the ambipolar electric field evolve alike those shown in Fig.~\ref{figure02} and Fig.~\ref{figure03}(a).

A magnetic field grows between the shock and the rarefaction wavefront, which are marked by the circles at $t=720$. The circles in Fig.~\ref{figure06} and in the forthcoming figures are placed at the same positions as in Fig.~\ref{figure02}. The magneto-waves escape from the density ramp and propagate into the dense- and dilute plasmas. The magnetic field starts to grow from noise levels at $t\approx 450$ and saturates at $t\approx 650$. Fitting an exponential curve to $B_\perp(x=102,t)$ gave the growth rate $\sigma_m = 0.12 \omega_p$ or $1.5\omega_{ce}$, which is slightly lower than the $\sigma_u=0.14 \omega_{p}$ measured in the simulation with $B_0=0$. Figures~\ref{figure06}(b,~c) evidences a correlation between both magnetic field components. A phase shift in space by 90$^\circ$ is observed near the black circle at $x=72$ and $t=1800$. A phase shift in time by 90$^\circ$ can be seen near the red circle at $x=132$ and $t=720$ for the wave that crosses the shock and propagates to increasing values of $x$. This phase shift characterizes circularly polarized Whistler waves. 

Whistler waves in cold electron plasma follow the dispersion relation Eqn.~\ref{eq1} provided that $\omega_{ce} < \omega_p$. In a 1D simulation, the magnetic amplitude along $x$ cannot change in space by $\nabla \cdot \mathbf{B} = 0$ and time (See Faraday's law in Eqn.~\ref{eq3}) and $\omega_{ce}$ is constant. The thermal ambient plasma is that with the lowest density and, hence, $\omega_p \gg \omega_{ce}$ for all $x$ in our simulation. In this case, the frequency range of Whistler waves is bounded by $\omega_{ce}$. Whistlers can leave the density ramp in both directions. If the Whistler waves keep their frequency $\omega$ unchanged, their wavenumbers $k$ must change as $k \propto \omega_{pe}$. Waves in the thermal ambient plasma should have a wavelength $2\pi/k$ that exceeds that in the thermal dense plasma by the factor $\sqrt{60}$. The wavelength of the magnetic oscillation near $x=0$ at $t\approx 10^3$ in Fig.~\ref{figure06}(b) is about 100, which gives the wavelength $\approx 800$ in the thermal ambient plasma and is close to what we observe in the simulation. 

\begin{figure}[ht]
\includegraphics[width=\columnwidth]{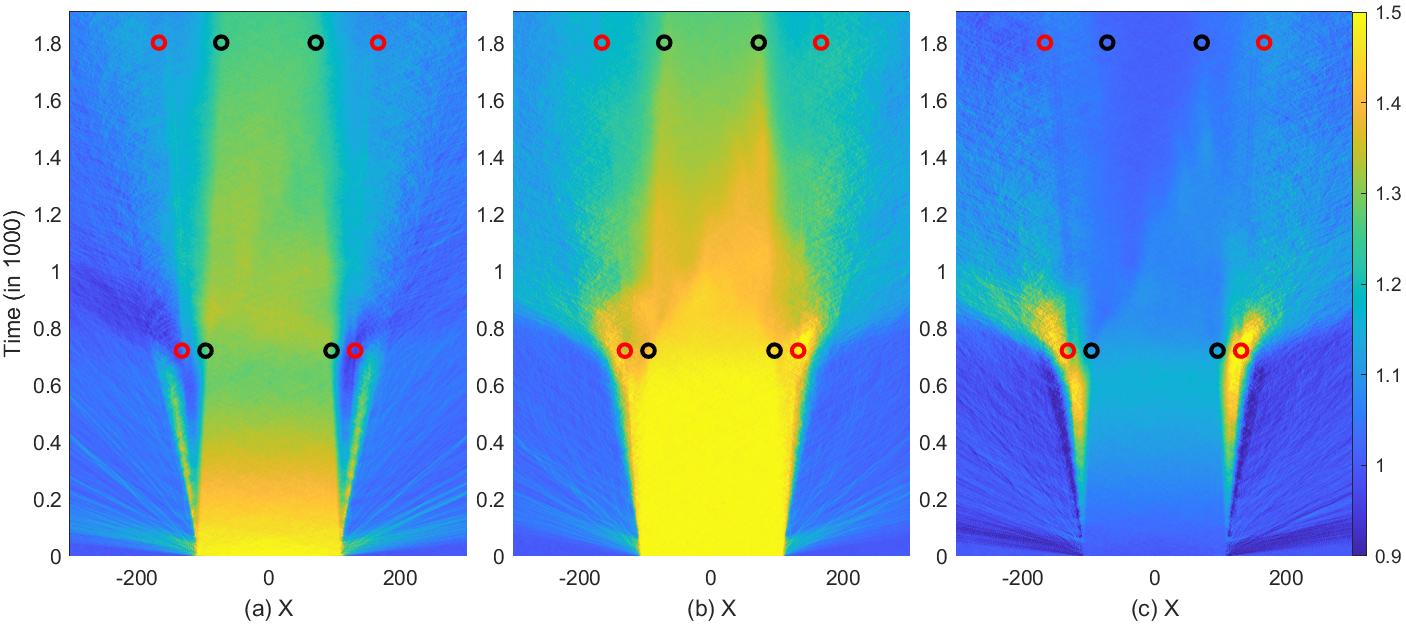}
\caption{The thermal energy $T_x(x)$ of the electrons is shown in panel~(a). Panels~(b) and~(c) show $T_\perp(x)=(T_y(x)+T_z(x))/2T_e$ and $A_\perp = (T_y(x)+T_z(x))/2T_x(x)$. All color scales are identical. The red circles show the locations of the shocks at $t=720$ and $t=1800$ while the black circles show those of the rarefaction wavefront.}
\label{figure07}
\end{figure}

Figure~\ref{figure07}(a) shows the thermal energy of electrons $T_x(x)$ along the background magnetic field. 
After a rapid initial decrease, the value of $T_x(x)$ stabilizes at a value that is 20\% higher than that in the ambient plasma. This temperature change is caused by the jump in the electric potential between the dense and the ambient plasma, which is mediated by the ambipolar electric field in Fig.~\ref{figure06}(a). Figure~\ref{figure07}(b) shows the thermal energy of electrons $T_\perp(x)=(T_y(x)+T_z(x))/2$ perpendicular to the background magnetic field. Its value in the dense plasma remains close to the initial one until $t=600$; the electron motions along and perpendicular to the background magnetic field are decoupled. After $t=600$, $T_\perp(x)$ rapidly decreases. The values $T_x(x)$ and $T_\perp (x)$ become similar after $t=1800$. Figure~\ref{figure07}(c) compares the perpendicular and parallel temperature by means of the thermal anisotropy $A_\perp (x) = (T_y(x)+T_z(x))/2T_x(x)$. Prior to $t\approx 600$, two maxima of $A_\perp(x)$ grow in the density ramps and downstream of both shocks. When the magneto-waves saturate at $t\approx 720$, the electrons between the shock and the rarefaction wavefront are heated along $x$ and cooled in the perpendicular direction. The thermal anisotropy in this interval decreases from 1.5 to 1.2.

The thermal anisotropy in Fig.~\ref{figure07}(c) above the black circle at $x=-96$ and $t=720$ decreases in the spatio-temporal interval, which is occupied by the magnetowave in Figs.~\ref{figure06}(b, ~c)  that propagates from the left density ramp into the thermal dense plasma. A less pronounced correlation between the magneto-wave and the change of $A_\perp$ can also be seen near the right density ramp. Equation~\ref{eq2} shows that the gyro-resonance between Whistler waves and electrons depends only on the parameters $\omega_{ce}$ and $v_{th,e}$. Figure~\ref{figure1} showed that Whistler waves in the thermal ambient plasma start to interact with electrons if their wavenumber $k > 0.02$ or $2\pi/k < 300$. Whistler waves in the thermal dense plasma in Fig.~\ref{figure06} have a much shorter wavelength than that. They interact with electrons and reduce $A_\perp (x,t)$ by scattering electrons in velocity space. 

\begin{figure}[ht]
\includegraphics[width=\columnwidth]{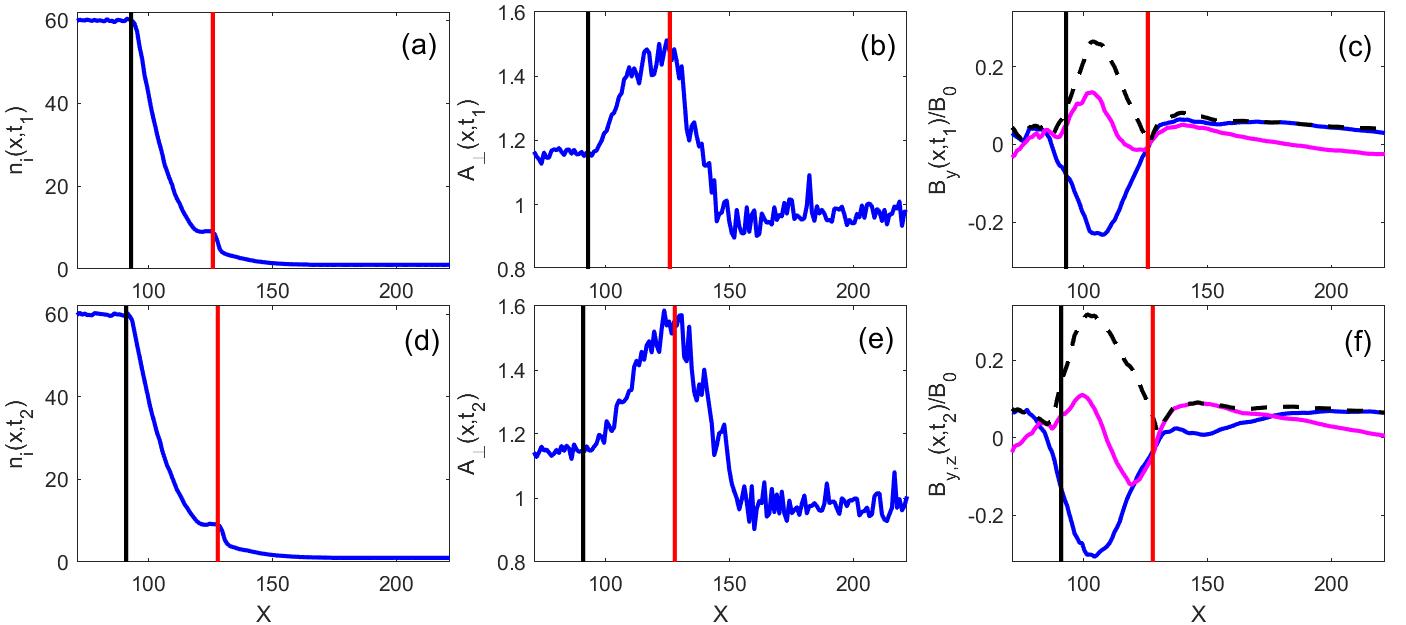}
\caption{Distributions of the ion density, thermal anisotropy, and magnetic field at the times $t_1=600$ (upper row) and $t_2=675$ (lower row): Panels (a,~d) plot the ion density $n_i(x)$ and~(b,~e) plot the thermal anisotropy $A_\perp(x)$. The magnetic field components $B_y(x)/B_0$ (blue) and $B_z(x)/B_0$ (magenta), and $(B_y^2+B_z^2)^{1/2}/B_0$ (dashed black curve) are plotted in panels (c,~f). The black (red) lines indicate the locations of the rarefaction wavefronts (shocks). These are 93 and 126 at $t_1$ and 91 and 128 at $t_2$.}
\label{figure08}
\end{figure}

Figure~\ref{figure08} compares $n_i(x)$, $A_\perp (x)$, $B_y(x)$, $B_z(x)$, and $(B_y^2(x)+B_z^2(x))^{1/2}$ at the times $t_1=600$ and $t_2=600 + \Delta_t$, where $\Delta_t=2\pi / \omega_{ce} \approx 75$. We normalize the magnetic field amplitude in this figure to $B_0$. 
Both ion density distributions show that the thermal dense plasma still maintains its original density over a broad interval. The thermal anisotropy $A_\perp(x)$ in this interval is about $1.15$. The ion density decreases from the value 60 at the rarefaction wavefront to about 8 when we enter the downstream region of the shock. This density value is higher than what we expect from the compression of the ambient plasma by the shock because the downstream plasma is not yet decoupled from the beam of accelerated ions. The density decreases from 8 to about 2 when we cross the shock. The shock reflects a significant fraction of ambient ions and we still find blast shell ions ahead of the shock as demonstrated by Fig.~\ref{figure02}(b). 
A gentle density decrease characterizes the foreshock. The peak value of $B_\perp/B_0=(B_y^2+B_z^2)^{1/2}/B_0$ is reached in the density ramp. It grows from $0.26$ at $t_1$ to $0.32$ at $t_2$. We find the strongest magnetic field to the left of the maximum value of $A_\perp(x)$; the exponential growth rate of the instability increases not only with the value of $A_\perp(x)$ but also with the local plasma frequency $\propto n_i(x)^{1/2}$. Like in the case of the Weibel-type modes, $B_\perp$ in Figs.~\ref{figure08}(c,~f) goes through 0 at the shock's position.

Let us estimate the frequency $\omega$ of the Whistler mode in the density ramp by means of Eqn.~\ref{eq1} assuming that it grows in a plasma with a uniform ion density and the local plasma frequency $\sqrt{30}\omega_p$. The wavenumber of the waves in Fig.~\ref{figure08}(c,~f) is $k_0 \approx 2\pi / 60$ where we normalized the wavenumber to the Debye length $\lambda_D=v_{th,e}/\omega_p$ of the thermal ambient plasma. Equation~\ref{eq1} predicts the wave frequency $\omega=0.45\omega_{ce}$ for $\omega_p/k_0c\approx 1.1$, which should give the phase change $0.9\pi$ between $t_1$ and $t_2$. Figures~\ref{figure08}(c) and Fig.~\ref{figure08}(f) show no change of the wave's phase during one electron gyro-period. Even if the wave frequency would be $\omega =0.45\omega_{ce}$, this would just be one-third of its exponential growth rate $\sigma_m$. The value $\sigma_m$ exceeds by far the growth rate of resonantly driven Whistler wave instabilities~\cite{Devine1995} and is comparable to $\sigma_u$. We conclude that the instability in the ramp is not a resonant Whistler instability. The magneto-waves in the density ramp must be driven by the same mechanism in the simulations with $B_0=0$ and $B_0=0.084$.

The magneto-wave in Fig.~\ref{figure05}(a) remained confined because an unmagnetized plasma does not support a propagating magneto-wave with $\omega \ll \omega_p$. This is no longer the case if $B_0 \neq 0$. The magneto-wave in the density ramp in Fig.~\ref{figure06} emits fast Whistler waves into the thermal ambient plasma and slow ones into the thermal dense plasma. If we keep $\omega/\omega_{ce}$ in Eqn.~\ref{eq1} fixed, the wavenumber scales as $k\propto \omega_p$. If we increase $\omega_p$ in Fig.~\ref{figure1}, the solution of the dispersion relation is pushed to the right, which decreases the phase velocity $\omega / k$ and the group velocity $\partial \omega / \partial k$. Increasing $\omega_p / \omega_{ce}$ slows down the waves. 

\begin{figure}[ht]
\includegraphics[width=\columnwidth]{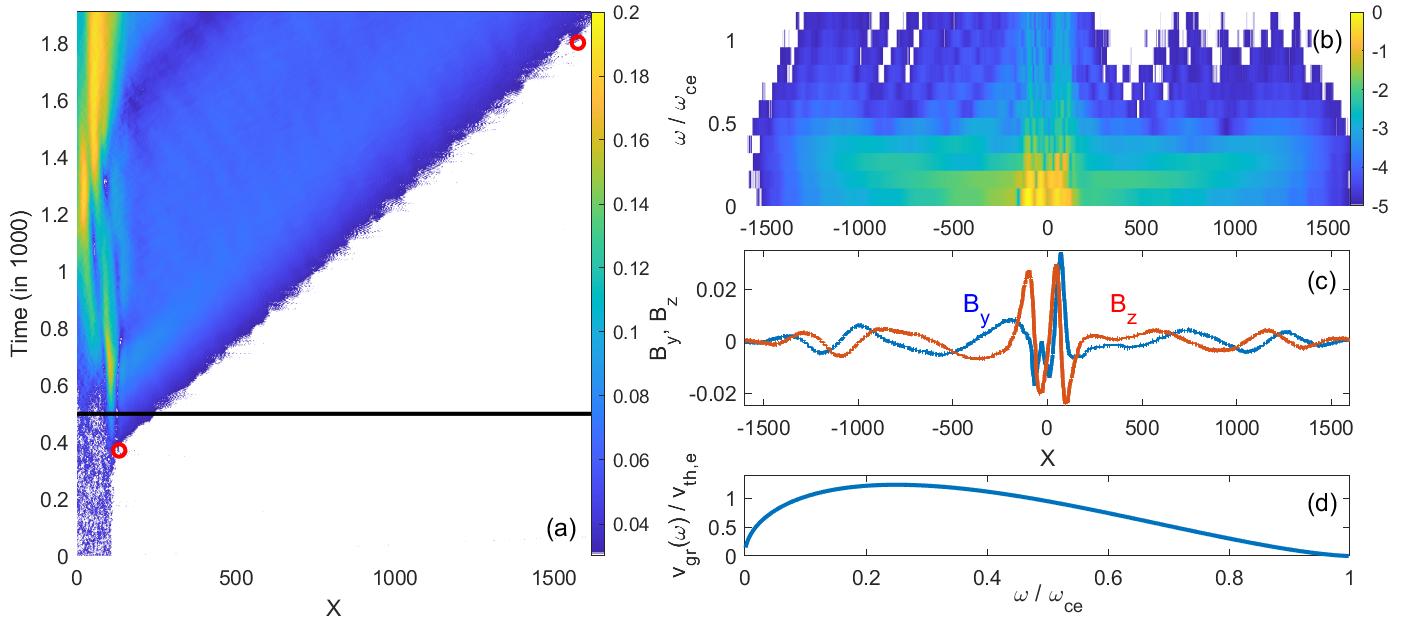}
\caption{The evolution of $B_\perp(x,t)^{1/2}$ is shown in~(a). The black horizontal line denotes $t=500$ and the red circles are placed at $(x,t)=(132,370)$ and $(1575,1800)$. Panel~(b) shows the spatial distribution of the wave frequency spectrum $P_B(x,\omega)$, which is integrated from $t=500$ to $t=1800$, normalized to its peak value, and displayed on a 10-logarithmic color scale. Panel~(c) shows the magnetic amplitudes at $t=1800$. Panel~(d) plots the group velocity of Whistler waves in cold plasma in units of the thermal speed of electrons with temperature $T_0$.}
\label{figure09}
\end{figure}

Figure~\ref{figure09}(a) shows the evolution of ${B_\perp(x,t)}^{1/2}$ in the thermal ambient plasma.
A coherent magnetic field grows at $x\approx 120$ at around $t=400$ in the density ramp. Whistlers propagate into the dense plasma and fill it. Another wavefront propagates at the speed $\approx v_{th,e}$ into the ambient plasma. Its amplitude remains constant. 

We compute the Fourier transform of $B_y(x,t)$ and $B_z(x,t)$ over $500\le t \le 1900$ and multiply them with their complex conjugate. We add up both power spectra to $P_B(x,\omega)=|B_y(x,\omega)|^2 + |B_z(x,\omega)|^2$ and show it in Fig.~\ref{figure09}(b). The wave power peaks in the spatial interval between the left- and right-moving shocks. This signal involves frequencies well below $\omega_{ce}$ and contains contributions of propagating Whistlers in the thermal dense plasma and of the aperiodically growing waves in the density ramp. The Whistler waves outside the interval, which is enclosed by both shocks, have a lower power and occupy the same frequency range as the waves in the thermal dense plasma. A frequency upshift of the power maximum to a frequency $\approx 0.3\omega_{ce}$ is visible near $|x| = 1000$. Figure~\ref{figure09}(c) plots $B_y(x,t)$ and $B_z(x,t)$ at the time $t=1800$. For $|x| > 500$, both magnetic field components have the phase shift $\approx 90^\circ$ that is typical for Whistler waves. Their peak amplitude far from their source is about $0.07B_0$. Figure~\ref{figure09}(d) plots the group velocity of Whistler waves as a function of $\omega/\omega_{ce}$. The peak speed is somewhat higher than $v_{th,e}$, which is the speed of the wave front in Fig.~\ref{figure09}(a). Its maximum is close to 0.25$\omega_{ce}$ and the frequency upshift of the Whistlers with increasing $x$ in Fig.~\ref{figure09}(b) is a selection effect; only the fastest waves can keep up with the wave front. 

\section{Discussion}

We have performed two one-dimensional simulations of a dense plasma, which expanded into an ambient plasma. At the simulation's start, both plasmas were spatially uniform and separated by a density jump. We selected initial conditions for the plasma, which could reproduce some aspects of the interaction between laser-generated blast shells and ionized residual gas. We modeled fully ionized nitrogen ions. Nitrogen gas is frequently used as the residual gas in laser-plasma experiments and turned into ambient plasma by secondary X-ray emissions from the laser-ablated target. Rarefaction waves propagated from both borders of the dense plasma into its interior and accelerated ions of the dense and ambient plasma in the opposite direction. The spatial interval, which was occupied by the dense plasma at the simulation's start, was wide enough so that the rarefaction waves were still separated by the time the simulation ended. Shocks at the fronts of the accelerated ions propagated into the ambient plasma with 1.5 times the ion acoustic speed. In time, a density ramp formed between the thermal dense plasma and the downstream plasma of the shock, which was characterized by a density that decreased and a mean velocity of the ions that increased with the distance from the rarefaction wavefront. The transfer of thermal energy from the electrons to the accelerating ions resulted in a thermal anisotropy of the electron distribution, which was strongest in the density ramp. One simulation modeled a plasma, which was unmagnetized at the simulation's start. In the second simulation, we aligned a background magnetic field with the simulation direction that had the amplitude $B_0$.

The simulation with no magnetic field confirmed previous studies that showed that the thermally anisotropic electron distribution, which develops in the density ramp, leads to an instability that is similar to that found by Weibel in spatially uniform plasma. We found that the magneto-waves were linearly polarized during the initial growth phase of the instability and that the magnetic field direction was set by noise. The growth of a linearly polarized magneto-wave restored the thermal isotropy of the electrons in the direction orthogonal to its magnetic field direction. At late times, we observed the growth of a second magneto-wave driven by the difference in the thermal energies of electrons along the wavevector and along the magnetic field of the initial magneto-wave. Although the magneto-waves grew in the density ramp and could not propagate away from it, diffusion let the magnetic field expand upstream of the shock. The magnetic amplitude went through a minimum at the shock position and the phases on both sides of the shock were phase-shifted by $\pi$. 

In the second simulation, the background magnetic field changed the spectrum of waves in the plasma. More specifically, it introduced Whistler waves. We selected a value for $B_0$ that gave an electron-cyclotron frequency $\omega_{ce}$ that was well below the electron plasma frequency $\omega_p$ everywhere. If $\omega_{ce} < \omega_p$, $\omega_{ce}$ is the maximum frequency accessible to Whistler waves. In this case, Whistlers can freely propagate from the dense into the ambient plasma without changing their frequency. A change in the plasma density changes only the wavelength of the Whistler waves. Our simulation showed that the value of $B_0$ was not large enough to change the instability mechanism in the density ramp. The oscillation frequency of the growing magneto-wave was small compared to its exponential growth rate and the latter was only marginally less than its counterpart in the simulation with no magnetic field. The thermal anisotropy of the electron distribution did not result in an instability that transferred energy from electrons to Whistler waves by means of resonant wave-particle interactions.

The presence of propagating electromagnetic wave modes with a frequency below $\omega_p$ implied that the magneto-wave did not remain confined to the density ramp. We observed Whistler waves, which propagated into the thermal dense plasma. Their short wavelength let the Whistlers resonate with electrons and they reduced the thermal anisotropy of the electrons in the thermal dense plasma, which had developed during the plasma expansion, by means of wave-particle interactions discussed elsewhere~\cite{Devine1995}. Whistlers also propagated from the density ramp into the ambient plasma. Their long wavelength in this low-density plasma implied that they could not interact resonantly with thermal electrons. They remained practically undamped. Their magnetic amplitude in the ambient plasma reached about 7\% of $B_0$, which might be high enough to be detectable in laser-plasma experiments. In a realistic setting, where the density gradient exceeds by far the one we could model in our simulation, their amplitude might be even higher than the one we observed here.

\section*{Acknowledgements}

The computations/data handling were enabled by resources provided by the National Academic Infrastructure for Supercomputing in Sweden (NAISS) at the National Supercomputer Centre partially funded by the Swedish Research Council through grant agreement no. 2022-06725.

\section*{Data availability statements}

All data that support the findings of this study are included within the article (and any supplementary files).

\end{document}